\documentclass[prl,twocolumn,showpacs,amsmath,amssymb,superscriptaddress,floatfix]{revtex4}
\usepackage[english]{babel}
\usepackage{graphicx}

\newcommand{\lsim}{\lesssim}
\newcommand{\gsim}{\gtrsim}

\begin{document}

\title{Kondo effect in triple quantum dots}

\author{R. \v{Z}itko}
\affiliation{J. Stefan Institute, Ljubljana, Slovenia}

\author{J. \surname{Bon\v ca}}
\affiliation{Faculty of Mathematics and Physics, University of
Ljubljana, Ljubljana, Slovenia}
\affiliation{J. Stefan Institute, Ljubljana, Slovenia}

\author{A. Ram\v{s}ak}
\affiliation{Faculty of Mathematics and Physics, University of
Ljubljana, Ljubljana, Slovenia}
\affiliation{J. Stefan Institute, Ljubljana, Slovenia}

\author{T. Rejec}
\affiliation{Faculty of Mathematics and Physics, University of
Ljubljana, Ljubljana, Slovenia}
\affiliation{J. Stefan Institute, Ljubljana, Slovenia}
\affiliation{Ben-Gurion University, Beer Sheva, Israel}

\date{\today}

\begin{abstract}
Numerical analysis of the simplest odd-numbered system of coupled quantum
dots reveals an interplay between magnetic ordering, charge fluctuations and
the tendency of itinerant electrons in the leads to screen magnetic moments.
The transition from local-moment to molecular-orbital behavior is visible in
the evolution of correlation functions as the inter-dot coupling is
increased. Resulting novel Kondo phases are presented in a phase diagram
which can be sampled by measuring the zero-bias conductance. We discuss the
origin of the even-odd effects by comparing with the double quantum dot. 
\end{abstract}


\maketitle

One of the main notions in molecular physics is the formation of extended
molecular orbitals when atoms are brought together to bind into a molecule.
Using atomic manipulation capabilities of modern low-temperature scanning
tunneling microscopes, the delocalization of electrons can now be directly
studied by adding individual atoms to a molecule-like chain of atoms on a
flat substrate \cite{wallis2002}. Alternatively, one can study the
transition from localized to extended behavior in the systems of coupled
quantum dots, which can also be considered as a variety of artificial
molecules. Unlike in atomic chains, bonding in coupled dots can be
continuously adjusted using the pinch gate electrodes \cite{hatano2004}.

The Kondo effect is a many-body phenomenon due to interaction between a
localized spin and free electrons. It leads to an increased conductance
through nanostructures and it was first observed in single quantum dots (QD)
\cite{goldhaber98}. The conductance as a function of temperature, gate and
bias voltages is in agreement with theoretical predictions that such dots
behave rather universally as single magnetic impurities \cite{pustilnik04}
and can be modelled using single impurity Anderson and Kondo models. By
analogy, the Kondo effect in multiple-dot systems can be described using
multiple-impurity Anderson model.

Measurements of the transport properties of multiple QD systems indicate
that the conductance is significantly different for even and odd number of
dots \cite{waugh95}. Double QD (DQD) systems are the simplest representation
of the two-impurity Anderson model. Their phase diagram features
single-impurity Kondo effect, two-electron Kondo effect and a
suppressed-conductance state due to anti-ferromagnetic (AFM) coupling
between the dots \cite{georges99,kikoin01}. Good understanding of these
different regimes was obtained using various numerical techniques
\cite{izumida00,lopez02,cornaglia05,bulka04}.

Much less is known about odd-number systems of coupled QD. Even the
properties of the triple quantum dot (TQD), the simplest member of the
family, have not been systematically resolved to date. There is clearly a
regime of Coulomb blockade at intermediate temperatures \cite{stafford94,
chen94}. Some properties of the Kondo regime at low temperatures are known
from previous studies using the perturbation theory \cite{oguri01}, the
numerical renormalization group \cite{hewson05}, scaling techniques
\cite{kuzmenko02}, the slave-boson mean-field theory \cite{jiang05} and
embedding \cite{busser04}. Some of these methods lead to conflicting results
for the same type of TQD, so further studies using reliable numerical
techniques are required.

\begin{figure}[htb]
\includegraphics[angle=0,width=8cm,scale=1.0]{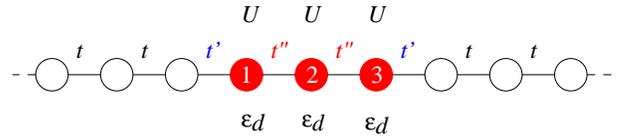} 
\caption{(Color online) The triple quantum dot embedded between two leads} 
\label{fig1}
\end{figure}

In this Letter, we study the TQD using two complementary numerical methods:
the constrained path Monte Carlo method (CPMC) \cite{zhang} and the
Gunnarsson-Sch\" onhammer variational method (GS) \cite{gunnarsson85}. The
conductance is calculated using the sine method \cite{rejec03}. We analyze
the low-temperature phases of the TQD embedded in series between two
metallic leads, Fig.~\ref{fig1}.

The TQD is modelled as a cluster of three equal QD described by Hamiltonians
$H_{i} = \sum_{\sigma} \epsilon_d n_{i\sigma} + U n_{i\uparrow}
n_{i\downarrow}$, $i=1,2,3$, where the on-site energy $\epsilon_d$ can be
varied using the gate voltage and $U$ is the electron-electron (e-e)
repulsion. The dots are interconnected with hopping matrix element $t''$ and
are symmetrically coupled to the leads with hopping matrix element $t'$. We
choose the chemical potential in the middle of the band, so that the model
is particle-hole (p-h) symmetric for $\epsilon_d+U/2=0$. 

For large $t''/U$, the system is in the {\it molecular-orbital Kondo regime}
when occupancy $\langle n \rangle = \sum_{i\sigma} \langle n_{i\sigma}
\rangle$ is odd. In the $U \to 0$ limit, the conductance through the TQD as
a function of the gate voltage exhibits three peaks corresponding to
resonant tunneling through distinct noninteracting levels: non-bonding,
bonding and anti-bonding {\it molecular-orbitals} with energies
$\epsilon_{\pm,0}=\pm \sqrt{2} t'',0$ and hybridizations
$\Gamma_{\pm,0}=\frac{3}{8}\Gamma, \frac{1}{2}\Gamma$, where
$\Gamma=2t'^2/t$. When interactions are switched on, the middle peak remains
at $\epsilon_d+U/2=0$, while the side-peaks are symmetrically shifted to
$\epsilon_d+U/2 \sim \pm (\sqrt{2}t''+ \frac{5}{16} U)+{\cal O}(U^2/t'')$ in
the molecular limit $\Gamma \to 0$. The $U$-term is a consequence of the
inter-orbital repulsion. The intra-orbital e-e repulsion is then given by
$U_\pm \sim \frac{3}{8} U + {\cal O}(U^2/t'')$ and $U_0 \sim \frac{1}{2} U +
{\cal O}[U^3/(t'')^2]$. The Kondo temperatures $T_{\pm, 0}$ for levels
$(\pm,0)$ can be determined from the Haldane formula \cite{haldane} for the
single QD using the corresponding parameters $U_{\pm,0}$.

\begin{figure}[htb]
\includegraphics[angle=0,width=7cm,scale=1.0]{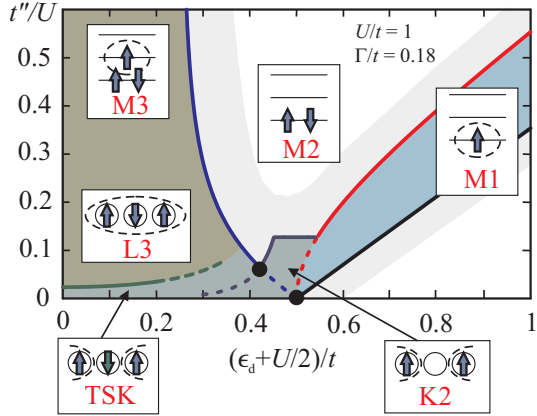}
\caption{(Color online)
M1, M3: molecular-orbital Kondo regime with $\langle n \rangle \sim 1,3$.
M2: non-conductive even-occupancy state.
L3: local Kondo regime with $\langle n \rangle \sim 3$.
TSK: two-stage Kondo regime.
K2: local double Kondo regime with $\langle n \rangle \sim 2$.
}
\label{fig2}
\end{figure}

The zero-temperature phase diagram of TQD exhibits several phases with
enhanced conductance, Fig.~\ref{fig2}. In the molecular-orbital Kondo regime
(shaded regions labelled 'M1' and 'M3') the conductance is $G/G_0 \gsim 0.5$
for low temperatures $T \ll T_{\pm, 0}$, while for intermediate temperatures
$T_{\pm,0} \ll T \ll \Gamma$ the system is in the Coulomb blockade regime
with $G \sim 0$ except along the border lines. Lightly shaded stripes of
width $\sim \Gamma/2$ represent the transition regions, where $G/G_0 \lsim
0.5$. Phase 'M2' is a non-conductive even-occupancy spin-zero phase, where two
electrons occupy the same molecular-orbital.

\begin{figure}[htb]
\includegraphics[angle=-90,width=8cm,scale=1.0]{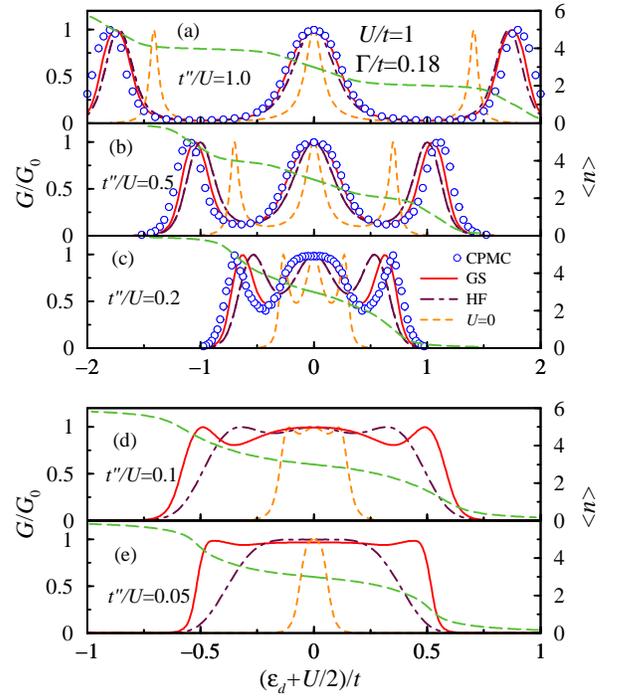}
\caption{(Color online) $G/G_0$ and $\langle n \rangle$
vs. $\epsilon_d+U/2$ (proportional to the gate voltage)
for various inter-dot hopping $t''$.
Note enhanced energy scale in panels (d) and (e).
Error bars of CPMC are smaller than the size of circles.
For comparison, we also show the Hartree-Fock (HF) as well as $U=0$ results.
}
\label{fig3}
\end{figure}

In Fig.~\ref{fig3} we present the zero-bias conductance $G$ along with the
total occupancy $\langle n \rangle$ as a function of the gate voltage for a
range of $t''$ and for fixed $U/t=1$ and $\Gamma/t=0.18$, calculated with
various methods as discussed below. For $t''/U\geq 0.2$, the system is in
the molecular-orbital regime. As the occupancy monotonically decreases from
6 (full TQD) to 0 (empty TQD), the conductance exhibits well resolved peaks
when occupancy is odd and valleys when occupancy is even \cite{goldhaber98}. 

As the TQD coupled to the leads is Fermi liquid \cite{hewson05}, the zero
temperature linear conductance is given by the sine formula derived in
Ref.~\onlinecite{rejec03}, $G = G_{0} \sin^{2} \left[ \frac{N}{4t}
(E_0-E_\pi) \right]$. Here $G_0 = 2e^2/h$ and $E_{0, \pi}$ are the ground
state energies of a large $N$-site auxiliary ring with embedded interacting
system, with periodic and anti-periodic boundary conditions, respectively.
Ground state properties are determined using CPMC and GS methods. 

In the CPMC method \cite{zhang}, the ground state wave function
$|\Psi_0\rangle$ is projected from a known trial function $|\Psi_T\rangle$
using a branching random walk that generates an over-complete space of
Slater determinants $|\phi\rangle$ and can be written as $|\Psi_0\rangle =
\sum_\phi c_{\phi} |\phi\rangle$, where $c_\phi>0$. To completely specify
$|\Psi_0\rangle$, only determinants satisfying $\langle\Psi_T|\phi\rangle>0$
are needed, because $|\Psi_0\rangle$ resides in either of two degenerate
halves of the Slater determinant space separated by a nodal plane. In this
manner, the minus-sign problem is alleviated. Extensive testing has
demonstrated a significant insensitivity of the results to reasonable
choices of $|\Psi_T\rangle$ \cite{zhang,Bonca}. The CPMC calculations were
performed on a ring of 100-180 sites. As the number of sites with
interaction is small, CPMC produces ground state energies with excellent
precision, typically of the order of $\Delta E/E=10^{-6}$.

In the GS method, the ground state energies are determined as the minimum
energy of the variational wave function $|\Psi\rangle =
\sum_{\lambda=1}^{64} c_{\lambda}P_{\lambda} \left|\tilde{0}\right\rangle$,
where $P_{\lambda}$ are the projection operators for the three sites of the
molecule, e.g., $P_1=\prod_{i,\sigma} (1-n_{i,\sigma})$, $P_2=n_{1,\uparrow}
(1-n_{1,\downarrow}) \prod_{i=2,3, \sigma} (1-n_{i,\sigma})$, \ldots,
$P_{64}=\prod_{i,\sigma} n_{i,\sigma}$. The vacuum state
$\left|\tilde{0}\right\rangle $ corresponds to the Fermi sea of a
non-interacting Hamiltonian $\tilde{H}$ with renormalized parameters. The
variational wave function is exact in the $U\to 0$ and $t'\to 0$ limits and
the test results for a single Anderson site perfectly match the exact Bethe
Ansatz results \cite{rejec03,comment,mravlje05}.

For $t''/U \geq 0.2$ the conductance obtained with CPMC and GS methods shows
good agreement. For lower $t''/U$ the CPMC method is no more applicable
since due to the computational restriction on the system size its energy
resolution is insufficient to describe the small Kondo scale
\cite{finitesize}. 

The {\it local regime} emerges when $t''/U$ decreases below $\sim 0.2$. The
orbital description breaks down as the orbitals start to overlap which leads
to qualitative changes in the system properties.  The relevant interaction
becomes $U$ rather than $U_{\pm,0}$. As presented in Fig.~\ref{fig3}(d,e),
$G/G_0 \sim 1$ near the symmetric point $\epsilon_d+U/2=0$, while in the
charge transfer region, $|\epsilon_d+U/2| \sim U/2$, the conductance
exhibits humps separated by dips. Due to larger $U$, the Kondo scale in the
local regime decreases. Note that for $t''/U \lsim 0.2$ the HF calculation
starts to deviate, signaling the onset of the strong coupling regime. For
even lower $t''/U$ the HF curves differ qualitatively from correct results.

The relation between the DQD and the TQD in the local regime is first
considered for $\langle n \rangle\sim 2$. This corresponds to the p-h
symmetric point of the DQD, where the two relevant energy scales are known
\cite{georges99} to be the exchange energy $J=4 (t'')^2/U$ and the Kondo
temperature $T_{1K}=T_K(U,\Gamma_1)$, with $\Gamma_1=\Gamma/2$, which is the
hybridization of a single impurity coupled to {\it one} lead \cite{haldane}.
In TQD, two electrons can reduce their energy by hopping within the TQD,
therefore instead of $J$ the relevant scale is the kinetic energy linear in
$t''$. The transition between the phase 'M2' to the phase 'K2' (see
Fig.~\ref{fig2}) occurs when $\sqrt{2}t'' \sim 2 T_{1K}$. The phase 'K2'
is a double Kondo phase, where the spin of each electron is screened by
electrons from the nearest lead. 

The behavior of the TQD near the p-h symmetric point, $\langle n
\rangle \sim 3$, is different from the behavior of the DQD near its
p-h symmetric point, $\langle n \rangle \sim 2$, due to the different
properties of integer and half-integer spin states. It is also
different from the behavior of the DQD near its charge-transfer
points, $\langle n \rangle \sim 1,3$, which exhibit Kondo effect only
for large $t''/U$. The TQD is fully conducting at the p-h symmetric
point for any $t''$ and as $t''/U$ is reduced the system goes
continuously through three different Kondo regimes.  The
molecular-orbital `M3' regime has already been discussed. For low
$t''/U$ there are two different regimes of local Kondo physics. (i) In
the {\it local Kondo} 'L3' {\it regime} for $J > 2 T_{1K}$, the TQD is
antiferromagnetically rigid and electrons form an ordered chain with
spin $1/2$ that exhibits the usual Kondo effect. Transition between
molecular-orbital 'M3' and local 'L3' is quite soft and determined by
the competition between kinetic ($t''$) and magnetic ($J$) scale. (ii)
In the {\it two-stage Kondo} 'TSK' {\it regime} for $J < 2 T_{1K}$,
each side QD ($i=1,3$) couples with $\Gamma_1$ to the nearest lead and
their spin is screened at $T \lesssim T_{1K}$, while due to very weak
coupling the remaining spin is screened at much lower temperature
$T_{2K} \propto T_{1K} \exp(-c T_{1K}/J)$ with $c \sim 1$
\cite{cornaglia05}.

In Table~\ref{table} we give quantitative relations for lines between
different phases of the system.  Boundaries between phases M1, M2, M3, and
L3 are determined in the molecular $\Gamma \to 0$ limit to the lowest
non-trivial order in $U$ (or $t''$). The expressions are in excellent
agreement with numerical results. We note that all transitions are smooth
(cross-overs) and there are no abrupt phase transitions.

\begin{table}[htbp!]
\centering
\begin{tabular}{@{}lll@{}}
\toprule
Phase 1 & Phase 2 & Condition \\
\colrule
empty & M1 & $\delta \sim U/2 + \sqrt{2}t''$ \\
M1 & M2 &
$\begin{cases} \delta
\sim U/2 + 3 (t'')^2 / U,
&  t'' \lesssim U \\
\delta \sim U/8 + \sqrt{2} t'',
&\ t'' \gtrsim  U
\end{cases}$ \\
M2 & M3, L3 &
$\begin{cases}
\delta  \sim U/2  - \sqrt{2} t'' + 3 (t'')^2/U,
&t'' \lesssim U \\
\delta \sim U \left(1/4 + \tfrac{3}{512} (U/t'')^2 \right),
& t'' \gtrsim  U
\end{cases}$ \\
M3 & L3 & $\sqrt{2}t'' \sim J $ \\
M2 & K2 & $\sqrt{2} t'' + 3 (t'')^2/U \sim 2 \min(T_{1K}, \Gamma_1)$ \\
L3 & TSK & $J \sim 2T_{1K}$ \\
\botrule
\end{tabular}
\caption{Definitions of boundaries and cross-over regions
between various phases in Fig.~\ref{fig2}.  Here
$\delta=\epsilon_d+U/2$ and $J=4(t'')^2/U$.
}
\label{table}
\end{table}

\begin{figure}[htb]
\includegraphics[angle=-90,width=8.0cm,scale=1.0]{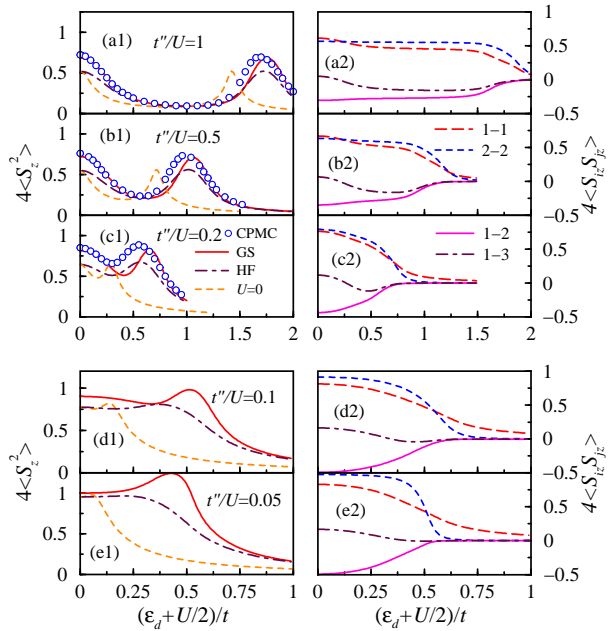}
\caption{(Color online)
Moments $\langle S_z^2 \rangle$ vs. $\epsilon_d+U/2$,
left panels (a1)-(e1).
Spin-spin correlation functions $\langle S_{iz}
S_{jz} \rangle$ for $(ij)=(11)$, $(22)$, $(12)$, and $(13)$ 
as calculated from the GS wave function,
right panels (a2)-(e2).
}
\label{fig4}
\end{figure}

The finger-print of the Kondo physics are not only the broadened
conductance peaks, but also the formation of a local moment in the
TQD, which is then screened by the electrons in the leads. In
Fig.~\ref{fig4} we present $\langle S_z^2 \rangle $, where $S_z =
\sum_{i=1}^3 S_{iz}$ is the z-component of the total spin of the TQD
(left panels), and local $i=j$, nearest neighbor $(ij)=(12)$ and
next-nearest neighbor $(ij)=(13)$ spin-spin correlations $\langle
S_{iz} S_{jz} \rangle$ (right panels).  For large $t''$, formation of
local moment (spin-doublet) is evident, but the spin is not saturated,
$4 \langle S_z^2\rangle < 1$.  For $t'' \to 0$ and $\langle n \rangle
\sim 2$, increased $4 \langle S_z^2\rangle > 1$ illustrates formation
of the 'K2' phase with large local moment and the absence of
inter-site spin correlations.  The two electrons are relatively
independent and therefore total spin is enhanced (note that for two
uncorrelated spins $4 \langle S_z^2 \rangle=2$). Each electron is
coupled by the Kondo mechanism to the adjacent lead. For $\langle n
\rangle \sim 3$, we see that neighboring spins tend to anti-align
while left-most and right-most spins align as $t''/U$ is reduced,
which represents formation of an antiferromagnetically ordered chain
when we enter the 'L3' phase.

Charge fluctuations $\Delta n^2=\langle (n-\langle n \rangle )^2\rangle$ and
the corresponding charge susceptibility $\chi_c=-(\pi \Gamma/4) \partial
\langle n \rangle/\partial \epsilon_d$ exhibit double peaks at positions of
the conductance peaks as an additional characteristic of the Kondo effect
\cite{mravlje05} and start to build a dip in the symmetric point for $t''/U
< 0.2$ (not shown here). This additionally signals the transition from
orbital to local Kondo regime and is in agreement with local spin formation
shown in Fig.~\ref{fig4}(d,e).

In summary, we have determined the phase diagram of the TQD. For strong
inter-dot coupling, the system behaves as an artificial molecule. The
extended "molecular orbitals'' are filled consecutively and the system
exhibits the usual Kondo effect when the number of confined electrons is
odd. For weak inter-dot coupling, local spin behavior is observed. The
cross-over from extended to local Kondo physics is illustrated by the smooth
evolution of the spin-spin correlation functions. In the vicinity of the p-h
symmetric point, the TQD tends to an antiferromagnetically ordered state
when the inter-dot coupling is decreased. For extremely weak coupling, the
Kondo screening of local moments occurs in two stages. In the charge
transfer regime the left and right sites tend to form two relatively
independent Kondo correlated states.

We expect that in the limit of extremely weak coupling, other complex phases
might exist in the immediate vicinity of the $\epsilon_d=0$ point, where
occupancy of the central dot changes abruptly. Furthermore, we believe that
the properties of all $n$ dot systems (with $n$ odd) are similar at the p-h
symmetric point: as the inter-dot coupling is decreased, the 
molecular-orbital Kondo regime $Mn$ is followed by an antiferromagnetically
locked Kondo regime $Ln$ and then by a two-stage Kondo regime.

We thank Y. Avishai, Y. Meir, and A. C. Hewson for helpful
discussions. Authors acknowledge support from the Ministry of
Higher education, Science and Technology of Slovenia under grant
Pl-0044 and from Ben-Gurion University.

\end{document}